\input cmtaco.sty
\font\fourteenbf=cmbx10 scaled \magstep2    
\magnification=\magstephalf

\vsize=9.2truein
\nohyphen
\mathfont\rm
\doublespace
\nopagenumbers
\headline={\ifnum\pageno>1\hss\tenrm --~\folio\tenrm ~-- \hss\else\hfil\fi}

\centerline{\fourteenbf THE HST KEY PROJECT ON}
\centerline{\fourteenbf THE EXTRAGALACTIC DISTANCE SCALE.}
\centerline{\fourteenbf XV. A CEPHEID DISTANCE TO THE FORNAX CLUSTER}
\vskip 0.1cm 
\centerline{\fourteenbf AND ITS IMPLICATIONS}
\vskip 1.5cm
\centerline{\bf BARRY F. MADORE$^{1}$, WENDY L. FREEDMAN$^{2}$, N. SILBERMANN$^{3}$}
\centerline{\bf PAUL HARDING$^{4}$,  JOHN HUCHRA$^{5}$, JEREMY R. MOULD$^{6}$}
\centerline{\bf JOHN A. GRAHAM$^{7}$, LAURA FERRARESE$^{8}$}
\centerline{\bf BRAD K. GIBSON$^{6,13}$, MINGSHENG HAN$^{9}$, JOHN G. HOESSEL$^{9}$}
\centerline{\bf SHAUN M. HUGHES$^{10}$, GARTH D. ILLINGWORTH$^{11}$, DAN KELSON$^{7}$}
\centerline{\bf RANDY PHELPS$^{2}$, SHOKO SAKAI$^{3}$, PETER STETSON$^{12}$}

\vskip 0.8cm
\vfill
\singlespace
\noindent
--------------------------------

\par\noindent
$1$  ~NASA/IPAC   Extragalactic   Database, Infrared   Processing  and
Analysis Center,

Jet Propulsion Laboratory, California Institute of Technology, MS~100-22,
Pasadena, CA ~91125
\par\noindent
$2$ ~Observatories of the Carnegie Institution of Washington,  813  Santa    Barbara  St.,  

Pasadena, CA ~91101
\par\noindent
$3$ ~Jet Propulsion Laboratory, California Institute of Technology,
MS~100-22, Pasadena, CA ~91125
\par\noindent
$4$  ~Steward Observatory,  University of Arizona,  Tucson, AZ ~85721
\par\noindent
$5$ ~Harvard Smithsonian Center   for Astrophysics,  60  Garden  Street,
Cambridge, MA ~02138
\par\noindent
$6$ ~Mt. Stromlo and Siding Spring Observatories,  
Institute of Advanced Studies,  Private Bag, 

Weston Creek Post Office, ACT 2611,  Australia
\par\noindent
$7$ ~Department  of Terrestrial   Magnetism, Carnegie  Institution  of
Washington, 

5241 Broad Branch Rd. N.W., Washington D.C. ~20015
\par\noindent
$8$ ~Hubble Felow, California Institute of Technology, MS~105-24 Robinson Lab, Pasadena, CA ~91125
\par\noindent
$9$ ~Department of Astronomy, University of  Wisconsin, 475 N. Charter
St., Madison, WI ~53706
\par\noindent
$10$ Royal Greenwich Observatory,  Madingley Road, Cambridge, UK ~CB3 OHA
\par\noindent
$11$ Lick   Observatory, University of  California, Santa  Cruz, CA
~95064  
\par\noindent
$12$ Dominion   Astrophysical  Observatory,  5071  W.    Saanich  Rd.,
Victoria, BC, Canada  ~V8X 4M6  
\par\noindent
$13$ Center  for Astrophysics  \& Space
Astronomy, University of Colorado, Boulder CO 80309-0389
\vfill\eject

\doublespace
\noindent
\centerline{\bf ABSTRACT}

{\bf Using the  Hubble Space Telescope (HST) thirty-seven  long-period
Cepheid  variables have been  discovered  in the Fornax Cluster spiral
galaxy NGC~1365  (Silbermann  et~al.  1999).   The resulting V  and  I
period-luminosity relations yield a  true distance modulus of $\mu_o =
$  31.35  $\pm$  0.07~mag,   which   corresponds to  a     distance of
18.6$\pm$0.6~Mpc.  This   measurement   provides  several routes   for
estimating the  Hubble Constant.  (1) Assuming  this  distance for the
Fornax  Cluster  as a   whole yields a  local   Hubble Constant of  70
($\pm$18)$_{random}$ [$\pm$7]$_{systematic}$ km/s/Mpc.     (2)    Nine
Cepheid-based distances to groups of galaxies out to and including the
Fornax and Virgo clusters  yield H$_o$ = 73 ($\pm$16)$_r$ [$\pm$7]$_s$
km/s/Mpc.  (3)  Recalibrating the I-band  Tully-Fisher relation  using
NGC~1365 and six nearby spiral galaxies, and applying  it to 15 galaxy
clusters  out  to 100~Mpc gives H$_o$   = 76 ($\pm$3)$_r$ [$\pm$8]$_s$
km/s/Mpc.   (4)   Using a broad-based    set  of differential  cluster
distance moduli ranging  from Fornax  to Abell 2147  gives  H$_o$ = 72
($\pm$3)$_r$  [$\pm$6]$_s$ km/s/Mpc.    And finally, (5)  Assuming the
NGC~1365 distance for the two additional Type  Ia supernovae in Fornax
and adding them  to the SNIa calibration  (correcting for  light curve
shape) gives  H$_o$ = 67  ($\pm$6)$_r$ [$\pm$7]$_s$ km/s/Mpc out  to a
distance in excess of 500~Mpc.  All five of these H$_o$ determinations
agree to within their statistical  errors.  The resulting estimate  of
the Hubble Constant combining all  of these determinations is H$_o$  =
72 ($\pm$5)$_r$ [$\pm$7]$_s$ km/s/Mpc.    An extensive tabulation   of
identified systematic  and statistical errors, and  their propagation,
is given.}

\vfill\eject
\centerline{\bf 1.~INTRODUCTION}
Hubble~(1929) announced his discovery of the expansion of the Universe
nearly  70 years ago.    Despite   decades of  effort, and   continued
improvements in  the  actual measurement of  extragalactic  distances,
convergence on a consistent value for the absolute expansion rate, the
Hubble constant, $H_o$,  has been elusive.   However, progress on  the
absolute calibration of the extragalactic  distance scale in the  last
few  years has  been rapid  and  dramatic (see, the recent proceedings
{\it ``The Extragalactic Distance Scale''} edited by Livio, Donahue \&
Panagia 1997 for instance,  containing  Freedman, Madore \&  Kennicutt
1997; Mould  et~al.   1997; Tammann \& Federspiel  1997;  and also see
Jacoby   et~al.  (1992) and  Riess,   Press  \& Kirshner  1996).  This
accelerated pace  has occurred primarily  as a  result of the improved
resolution of the Hubble    Space Telescope (HST) and  its  consequent
ability to discover classical Cepheid variables  at distances a factor
of ten further than  can routinely be  achieved from the ground.  As a
result,  accurate zero points to  a number of recently refined methods
which can measure precise relative  distances beyond the realm of  the
Cepheids have become  available.  These combined efforts are providing
a more accurate distance scale for local  galaxies, and are indicating
a convergence   among   various  secondary   distance  indicators   in
establishing an absolute calibration of the far-field Hubble flow.

The discovery of Cepheids with HST has proven to be very efficient out
to and even somewhat beyond distances of $\sim$20~Mpc.  Soon after the
December 1993 HST servicing mission the measurement of Cepheids in the
Virgo  cluster (part of  the  original  design specifications for  the
telescope)  became feasible~(Freedman et~al.   1994a).  The subsequent
discovery   of  Cepheids  in  the  Virgo  galaxy M100~(Freedman et~al.
1994b;  Ferrarese  et~al.  1996) was an   important  step in resolving
outstanding differences  in  the  extragalactic  distance scale (Mould
et~al.  1995).  The Virgo cluster is complex both in its geometric and
its kinematic structure, and there still remain large uncertainties in
both  the  velocity and  distance to  this cluster.   Hence, the Virgo
cluster is not an ideal test  site for an unambiguous determination of
the cosmological expansion  rate   or  the calibration  of   secondary
distance  indicators.  In this paper  we discuss the implications of a
Cepheid distance to the next major cluster  of galaxies, Fornax, which
is a  simpler system than Virgo.

In the companion paper to this one (Silbermann et~al. 1999) we present
the Cepheid photometry and PL relations for  the Cepheids in NGC~1365.
In  Madore et~al.  (1998)  we briefly  discussed the determination  of
$H_o$ based  on the distance of NGC~1365   and the Fornax  Cluster, in
addition to a calibration of a local Hubble  expansion-rate plot.  The
Fornax   cluster is comparable in distance   to  the Virgo cluster~(de
Vaucouleurs 1975),  but it is found  almost  opposite to  Virgo in the
skies of the southern hemisphere.  The Fornax  cluster is less rich in
galaxies  than Virgo~(Ferguson \&  Sandage    1988), but it is    also
substantially more compact  than its northern counterpart  (Figure~1).
As  a result of  its lower mass, the influence  of Fornax on the local
velocity field is less dramatic than  that of the  Virgo cluster.  And
because of its compact nature, questions concerning the membership and
location in the cluster of individual  galaxies are significantly less
problematic; the   back-to-front geometry   is far simpler    and less
controversial    than that  of    the Virgo  cluster.  Clearly, Fornax
provides   a  much more  interesting  site  for  a   test of the local
expansion rate.

In   the context  of  the Key   Project on  the Extragalactic Distance
Scale~(Kennicutt, Freedman \& Mould 1995), there are several important
reasons to secure a distance to the Fornax cluster. The Fornax cluster
serves as  both   a probe of  the  local  velocity field  and  a major
jumping-off point for several secondary distance indicators, which can
be used to probe a  volume of space  at least 1,000  times larger.  To
obtain a distance to the Fornax cluster, the  H$_0$ Key Project sample
includes three member galaxies; the first of these, discussed here, is
the Seyfert~1  galaxy NGC~1365,  a  striking, two-armed, barred-spiral
galaxy with   an active galactic  nucleus.  Two   additional galaxies,
NGC~1425 and NGC~1326A, have also been imaged  with HST and those data
are being processed; preliminary reduction shows  that the distance to
NGC~1326A   (Prosser   et~al., in   preparation)    lies  within the
uncertainties quoted here for NGC~1365.

\vfill\eject
\medskip
\centerline{\bf  2. ~NGC~1365 AND THE FORNAX CLUSTER}
\medskip

Three lines  of  evidence independently  suggest  that  NGC~1365 is  a
representative,   physical  member  of  the   Fornax cluster.   First,
NGC~1365 is  almost directly along  our line  of  sight to Fornax: the
galaxy is projected only $\sim$70 arcmin (380  kpc) from the geometric
center  of  the   cluster, whereas  the    radius of  the  cluster  is
$\sim$100~arcmin~(540 kpc; Ferguson 1989,  and see also Figure~1).  In
addition,  NGC~1365 is also   coincident  with the  Fornax cluster  in
velocity space.  The observed  velocity of NGC~1365 (+1,636~km/sec) is
only +234~km/sec larger than the cluster mean,  and is well inside the
cluster velocity dispersion  (see below.)  Finally,  we  note that for
its rotational velocity, NGC~1365 sits  within 0.02~mag of the central
ridge line  of the apparent  Tully-Fisher  relation relative  to other
cluster  members  defined by   recent  studies of   the Fornax cluster
(Bureau, Mould \& Staveley-Smith 1996; Schroder 1995).

\noindent
NGC~1365 is large in angular  size, and it  is very bright in apparent
luminosity as compared  to any other  galaxy in the immediate vicinity
of the Fornax cluster.  One might question whether  on this basis, NGC
1365   is a true member   of the Fornax   cluster.   Correcting for an
inclination  of 44\deg,~(see  Bureau et~al.   1996)  the 21cm  neutral
hydrogen   line    width   of   NGC~1365       is    found  to      be
$\sim$575~km/sec~(Bureau  et~al.   1996; Mathewson, Ford  \&  Buchhorn
1992).  A   $\pm5\deg$ error in this determination   would result in a
10\%  uncertainty in the  derived line  width.  Using the Tully-Fisher
relation as a {\it  relative} guide to  intrinsic size and luminosity,
this rotation rate places NGC~1365 among the most luminous galaxies in
the   local Universe;  brighter  than  M31 or  M81, and  comparable to
NGC~4501 in  the Virgo cluster or  NGC~3992 in the Ursa Major cluster.
Thus, the Tully-Fisher relation predicts that  NGC~1365 is expected to
be apparently bright, even at the distance  of the Fornax cluster, and
that its observed global properties  are consistent with membership in
that cluster.

\vfill\eject
\medskip
\medskip
\medskip
\medskip
\medskip
\medskip
\centerline{\bf  3. ~THE MEAN VELOCITY AND VELOCITY DISPERSION OF FORNAX}

 The systemic (heliocentric) velocity  and velocity dispersion  of the
main population of  galaxies in Fornax  are well defined.  A search of
the      NASA/IPAC         Extragalactic          Database       (NED:
http://nedwww.ipac.caltech.edu,  version   release  date 01/98)    for
galaxies within   6\deg~  of  the  Fornax  cluster center   and having
published  redshifts  $\le$2,500~km/sec   produced a   sample  of  106
galaxies; this was  was then supplemented  with 4 additional redshifts
from ZCAT, (Huchra, Geller, Clemens,  Tokarz \& Michel 1992; the  1998
edition of ZCAT is  available via anonymous ftp from fang.harvard.edu)
and   7 recently published dwarf   galaxy redshifts from Drinkwater \&
Gregg (1998), giving  a total of  117 redshifts.  The  distribution of
these 117 objects projected  on the sky is  shown in Figure~2; and two
`pie   diagrams'   illustrating     the   sample  distribution      in
position-velocity  space  are  shown  in    Figure~3.   In all   three
representations, ellipticals are shown  as filled circles, spirals  as
open circles.  While the core of the cluster is demonstrably dominated
by E/S0 galaxies, there  is no other   obvious segregation of  the two
populations:   spirals and ellipticals   being  coincident and largely
co-spatial.   After subdividing  the sample by  morphological type, 39
spirals/irregular galaxies give V = 1,399~km/sec and $\sigma =
\pm334$~km/sec, 78~E/SO galaxies give  V = 1,463~km/sec with $\sigma =
\pm347$~km/sec.  The mean velocity of the spirals agrees with the mean
for the ellipticals to within 0.2 $\times
\sigma$,  the velocity  dispersion of the   system. The combined sample 
of 117 galaxies has an unweighted mean of V = 1,441~km/sec and $\sigma
=$ $\pm342$~km/sec which we  adopt hereafter (see also Schroder  1995;
Han \& Mould 1990.)  The  velocity off-set of +195~km/sec for NGC~1365
with respect to this  mean is less than 2/3   of the cluster  velocity
dispersion.  \footnote  {$^1$}{Given that the  redshifts  are of mixed
quality with regard to reported uncertainties,  differing by up to two
orders of  magnitude,  we also  calculated the systemic
velocity of the  cluster  weighting the  individual  velocities by the
inverse   square of the   internal errors.   That  solution  gives V =
1,405~km/sec, agreeing with the  adopted value to within 3\%,  despite
the fact that it   is heavily weighted  by  only  a dozen or  so  high
precision points in the distribution (see Figure 4).}

\vfill\eject

\medskip
\medskip
\centerline{\bf  4. ~HST OBSERVATIONS AND THE CEPHEIDS  IN NGC~1365}  
\medskip

Using the  {\it Wide Field  and  Planetary Camera 2}  on  HST, we have
obtained  a set of  12-epoch observations of  NGC~1365.  The observing
window of 44 days, beginning August 6,  and continuing until September
24,  1995, was  selected    to  maximize target  visibility,   without
necessitating any roll of the targeted field of view.  Sampling within
the window  was prescribed  by a  power-law distribution,  tailored to
optimally   cover the   light  and  color   curves   of Cepheids  with
anticipated periods in the range 10 to 60 days (see (3) for additional
details).  Contiguous with  4 of the 12  V-band epochs (5,100~sec each
through the  F555W filter),  I-band exposures (5,400~sec  each through
the F814W filter) were also obtained so as to allow a determination of
reddening corrections for the Cepheids.

All  frames  were pipeline pre-processed  at  the {\it Space Telescope
Science Institute}  in  Baltimore and subsequently analyzed  using two
stellar photometry   packages,  ALLFRAME  (Stetson 1994)   and  DoPhot
(Schecter et~al.  1993), in  order   to quantify potential  systematic
differences  in the  two  reduction programs.  Zero-point calibrations
for the photometry were adopted from Holtzmann, J.  et~al.  (1995) and
Hill et~al.  (1998),  which agree to  0.05~mag on average.  Details on
the DoPhot and ALLFRAME reduction  and analysis of  this data set  are
presented elsewhere~(Silbermann  et~al.  1999).  We are also currently
undertaking artificial  star tests on  these   frames to quantify  the
uncertainty due to crowding (Ferrarese et~al., in preparation).

Detailed  information on  the    52 Cepheid candidates   discovered in
NGC~1365 can be found in Silbermann et~al.  (1999). The phase coverage
in all  cases is sufficiently dense  and uniform that  the form of the
light curves is clearly delineated.  We have adopted a sample of 37 of
these variables as  being  unambiguously  classified  as  high-quality
Cepheids  on   the basis of  their  distinctively   rapid brightening,
followed by a  long,  linear decline phase  (for  both the  DoPhot and
ALLFRAME   variable-star  candidates).     Periods,  obtained using  a
modified Lafler-Kinman   algorithm~   (Lafler  \&   Kinman  1965), are
statistically  good  to   a  few percent,   although  in   some  cases
ambiguities larger than this do  exist as a  consequence of the narrow
observing window and the restricted number of cycles (between 1 and 5)
covered within the 44-day window. A variety of other samples, selected
and reduced in a number of different  ways are discussed in Silbermann
et~al. To  the dgree  that the  distances  all agree  to within  their
quoted errors the broad conclusions resulting from  this paper are not
affected by choice of sample.

The  resulting V and I  period-luminosity relations for the select set
of 37 Cepheids (using  intensity-averaged magnitudes) are shown in the
upper and lower panels of Figure~5, respectively.  This sample differs
slightly from  that adopted in Silbermann  et~al.  (1999)  only in the
fact that the three 50-day Cepheids are retained in this analysis. The
derived apparent  moduli   are  $\mu_V =$   31.68$\pm(0.05)_r$~mag and
$\mu_I  =$  31.55$\pm(0.05)_r$~mag.   Correcting  for  a derived total
line-of-sight reddening of $E(V-I)_{N1365} =  $ 0.14~mag (derived from
the  Cepheids themselves) gives a  true  distance modulus of $\mu_0 =$
31.35$\pm(0.07)_r$~mag.  This corresponds to a distance to NGC~1365 of
18.6$\pm(0.6)_r$~Mpc,  which is within 2\%   of the value derived from
phase-weighted  magnitudes of the  slightly smaller Cepheid dataset as
given  in  Silbermann et~al.   The quoted error   at  this step in the
discussion  quantifies  only  the   statistical  (random)  uncertainty
generated by photometric errors in the ALLFRAME data combined with the
intrinsic magnitude and color width of the Cepheid instability strip.

Extensive reviews of the distance to the Fornax cluster (especially in
the  context  of a differential  comparison  with the  distance to the
Virgo cluster) can be found in two recent publications (Bureau, Mould
\& Staveley-Smith 1996, Table 3; Schroder 1996, Table 6.1). The former 
authors  quote a distance of  16.6   $\pm$ 3.4~Mpc  which in turn  was
consistent with  a value of 16.9   $\pm$ 1.1~Mpc reported  in the same
year by McMillan, Ciardullo \& Jacoby (1996).

\vfill\eject
\medskip
\medskip
\centerline{\bf  5. THE HUBBLE CONSTANT} 

We now discuss the impact of a Cepheid  distance to the Fornax cluster
in estimating the Hubble constant.  Before doing so we must make clear
the limited   context  and focussed nature   of  this  paper.   We are
interested in exploring the consequences of Cepheid-based distances in
general and the impact of a Cepheid distance to Fornax in specific, on
the  determinations of the  extragactic distance scale (and the Hubble
constant) directly dependent upon the Cepheids.   This is not intended
to be a review   of all measures of   the Hubble constant.  Nor  do we
revisit methods  that do not penetrate  the flow any further  than the
Cepheids  themselves  now do.   At the   time of writing,  this latter
exclusion applied to  the planetary nebula  luminosity function (PLNF)
method and to the surface brightness fluctuation (SBF) method, neither
of  which (see  Jacoby et~al.   1992  for an extensive discussion  and
review)  extended further than   Fornax, the  subject of this  Cepheid
paper.  In  the mean time HST observations  by  Lauer, Tonry, Postman,
Ajhar \& Holtzmann (1998) and by Jensen, Tonry  \& Luppino (1998) have
extended the  SBF method out to  the far field,  and they  determine a
values  of $H_o$  = 89 and   87 $\pm$ 10~km/sec/Mpc.  A  comprehensive
re-analysis of  SBF  and other methods will  be  presented in a  later
series Key Project team papers. This is an interim report.

Below we present and   discuss several  independent estimates of   the
local expansion  rate, where the   analysis is based  both on  the new
Fornax   distance and the  distances  to  other Key Project  galaxies,
consistently scaled  to a true  distance modulus of 18.50~mag (50 kpc)
for the   Large  Magellanic Cloud.   At the   end we intercompare  the
results for convergence and consistency.   The first estimate is based
solely   on the Fornax  cluster, its   velocity  and the Cepheid-based
distance to one of its members.  It samples the flow in one particular
direction  at a distance of $\sim$20~Mpc.   We  then examine the inner
volume of space, leading up to and including both the Virgo and Fornax
clusters.  This  has the added  advantage  of averaging over different
samples and a variety of directions, but it is still limited in volume
(to  an average distance of  $\sim$10 Mpc), and  it  is subject to the
usual caveats concerning bulk flows  and the adopted Virgocentric flow
model (Table~1).  The third  estimate   comes from using the   Cepheid
distance to Fornax to lock into secondary distance indicators, thereby
allowing us   to step  out to  cosmologically   significant velocities
(10,000~km/sec and  beyond)  corresponding to  distances  greater than
100~Mpc.   Averaging over the sky,   and working  at large  redshifts,
alleviates  the  flow  problems.   Examining  consistency  between the
independent  secondary  distance  estimates,  and then  averaging over
their far-field estimates should  provide a more systematically secure
value of $H_o$ and, more importantly, a measure of its external error.
Comparison of  the   three `regional'  estimates  (Fornax,  local  and
far-field) then  can be  used to provide   a check on  the systematics
resulting from  the   various assumptions made  independently  at each
step.

\medskip
\medskip
\centerline{\bf  6. UNCERTAINTIES IN THE FORNAX  CLUSTER DISTANCE AND VELOCITY}

The two panels of Figure~1  show a comparison  of the Virgo and Fornax
clusters of galaxies drawn to scale, as seen projected on the sky. The
comparison of  apparent   sizes  is  appropriate given that   the  two
clusters  are  at approximately the   same distance from   us.  In the
extensive Virgo cluster   (right panel), the galaxy  M100  can be seen
marked  $\sim$4\deg~ to the   north-west of the elliptical-galaxy-rich
core;  this corresponds to an impact  parameter  of 1.3~Mpc, or 8\% of
the distance from  the LG  to the Virgo  cluster.  The  Fornax cluster
(left panel) is  more centrally concentrated  than Virgo, so that  the
back-to-front    uncertainty  associated  with   its three-dimensional
spatial extent is reduced for  any randomly selected member.   Roughly
speaking, converting the total angular  extent of  the cluster on  the
sky ($\sim$3\deg~  in  diameter;  Ferguson \&  Sandage 1988)   into  a
back-to-front  extent, the error   associated with any randomly chosen
galaxy   in the   Fornax  cluster,   translates  into a   few  percent
uncertainty in  distance;  this uncertainty in  distance can be
reduced when more distances to spirals in Fornax have been measured.

Here, we note that the infall-velocity correction  for the Local Group
motion    with respect to   the   Virgo  cluster (and  its  associated
uncertainty) becomes a minor  issue for the  Fornax cluster.  This  is
the result of a fortuitous combination of geometry and kinematics.  We
now have Cepheid distances from the Local Group to both the Fornax and
Virgo clusters.   Combined with their  angular separation  on the sky,
this immediately  leads  to the physical   separation between the  two
clusters.   Under the assumption that  the Virgo cluster dominates the
local velocity  perturbation field  at the  Local Group {\bf  and}  at
Fornax, we can calculate the velocity perturbation at Fornax (assuming
that  the flow     field  amplitude scales   with    $1/R_{Virgo}$ and
characterized by  a $R^{-2}$  density distribution,   Schechter 1980).
>From  this  we then  derive the   flow  contribution to   the measured
line-of-sight radial velocity, as seen from the Local Group.  Figure~6
shows the distance scale structure (left panel) and the velocity-field
geometry (right panel) of  the Local Group--Virgo--Fornax system.   An
infall   velocity of the Local  Group  toward Virgo  of +200~km/sec is
obtained  by minimizing the velocity  residuals  for the galaxies with
Cepheid-based distances.  This value   is in good agreement with  that
estimated by Han \& Mould  (1990).  We adopt 200$\pm$100~km/sec, which
results in a projected   Virgocentric  flow correction for   Fornax of
--45$\pm$23~km/sec.

\medskip
\medskip
\centerline{\bf  7. $H_o$ AT  FORNAX, AND  ITS   UNCERTAINTIES}  

Correcting to the barycentre of the Local Group ($-$90~km/sec) and for
the $-$45~km/sec component of  the Virgocentric flow derived above, we
calculate  that  the cosmological  expansion   velocity of  Fornax  is
1,306~km/sec.  Using our Cepheid distance of 18.6~Mpc for Fornax gives
$H_o =  70~(\pm18)_r~~[\pm7]_s$~km/sec/Mpc.  The first uncertainty (in
parentheses) includes random errors  in the distance derived from  the
PL fit to the  Cepheid data (see Table  1), as well as random velocity
errors in the adopted  Virgocentric  flow, combined with  the distance
uncertainties to Virgo propagated through  the flow model.  The second
uncertainty (in square brackets) quantifies the currently identifiable
systematic errors associated with the adopted mean velocity of Fornax,
and the adopted zero point of the PL relation (combining in quadrature
the LMC distance error, a measure  of the metallicity uncertainty, and
a conservative  estimate of the  stellar photometry errors).  Finally,
we note that according to the Han-Mould model~(Han \& Mould 1990), the
so-called  ``Local Anomaly'' gives the Local   Group an extra velocity
component of approximately $+$73~km/sec towards Fornax.  If we were to
add that correction  to our local estimate,  the Hubble constant would
increase to $H_o$ = 74~km/sec/Mpc.

Given  the highly  clumped   nature of   the local   universe  and the
existence  of   large-scale  streaming velocities,   there is  still a
lingering  uncertainty about the total  peculiar  motion of the Fornax
cluster with  respect to  the cosmic  microwave  background restframe.
Observations of flows, and the determination of the absolute motion of
the Milky Way  with respect to the  background  radiation suggest that
line-of sight velocities $\sim$300~km/sec are not uncommon~({\it e.g.}
Coles \& Lucchin 1995   and references therein).  The  uncertainty  in
absolute motion of Fornax with respect to the Local Group then becomes
the  largest outstanding  uncertainty   at this   point in  our  error
analysis:  a  300~km/sec flow velocity for  Fornax  would  result in a
systematic error in the Hubble constant of $\sim$20\%.  We can revisit
this issue, however, following an  analysis  of more distant  galaxies
made later in this section.

\medskip
\medskip
\centerline{\bf  8. THE  NEARBY FLOW   FIELD}  

We now step back  somewhat and investigate the  Hubble flow between us
and  Fornax,   derived from galaxies  and   groups  of galaxies inside
20~Mpc, each having Cepheid-based  distances and  expansion velocities
individually   corrected   for    a   Virgocentric  flow   model  (see
Kraan-Korteweg  1986,  for example).   These   data are  presented  in
Figure~7.  At 3~Mpc the M81-NGC~2403 Group (for which both galaxies of
this  pair    have Cepheid distance   determinations)   gives  $H_o =$
75~km/sec/Mpc after  averaging their two velocities.  \footnote {$^2$}
{If  instead we define  the M81-NGC~2403 group  velocity by the simple
average  of  16 members (falling within  5   degrees of the primaries,
having  radial velocities  in   NED less  than  +350~km/sec) then  the
calculated Hubble constant for  that group {\it  increases} by 20\% to
90~km/sec/Mpc.  Other   nearby galaxies   with Cepheid distances   are
problematic: several, like NGC~3109 are borderline Local Group members
whose   velocities  are undoubtedly  dominated  by  M31  and the Milky
Way. NGC~300 and other members of the South Polar (Sculptor) Group are
appreciably strung out  (1.7 to    4.4~Mpc  along our  line   of  sight
according  to Jerjen, Freeman   \&   Binggeli 1998) and so   averaging
velocities makes   no   sense until  Cepheid  distances   to the other
(significant) members of the  group are obtained. Similar reasons were
invoked for omitting NGC~5253 until such time as  other members of the
extended M83 Group have Cepheid  distances.}   Working further out  to
M101, the NGC~1023 Group and  the Leo Group,  the calculated values of
$H_o$ range from 62 to 99~km/sec/Mpc.  An average of these independent
determinations   including  Virgo   and    Fornax,  gives   $H_o    =$
73~$(\pm16)_r$~km/sec/Mpc, where flow  uncertainties are  added to the
random error estimate  for later intercomparisons of  Hubble constants
derived  from independent  methods    and  volumes of   space.    This
determination,  as  before,  uses  a Virgocentric flow   model  with a
$1/R_{Virgo}$ infall velocity fall-off, scaled to a Local Group infall
velocity of +200~km/sec.

The foregoing  determination  of  $H_o$  is again predicated   on  the
assumption that the  infall  flow-corrected velocities of  both Fornax
and Virgo  are not further perturbed by  other  mass concentrations or
large-scale  flows,  and  that the   25,000  Mpc$^3$  volume  of space
delineated by   them is at rest  with  respect to  the  distant galaxy
frame.  To avoid these local uncertainties we now step out from Fornax
to  the distant flow field.  There  we explore three applications: (i)
Use of  the  Tully-Fisher  relation calibrated by    published Cepheid
distances locally, and now  including   NGC~1365 and about  two  dozen
additional galaxies     in  the Fornax     cluster.   Ultimately these
calibrators  are tied  into  the distant flow  field  at 10,000~km/sec
defined by   the    the    Tully-Fisher   sample of     galaxies    in
clusters~(Aaronson et~al.  1980;  Han 1992).  (ii) Using  the distance
to Fornax to tie into averages  over previously published differential
moduli for  independently   selected  distant-field   clusters,  (iii)
Recalibrating the Type Ia supernova luminosities at maximum light, and
applying that calibration to events as distant as 30,000~km/sec.

\vfill\eject
\medskip
\medskip
\centerline{\bf  9. BEYOND FORNAX:  THE TULLY-FISHER RELATION}  

Quite  independent of  its association  with  the Fornax  cluster as a
whole,  NGC~1365 provides   an  important calibration  point   for the
Tully-Fisher relation  which  links the   (distance-independent)  peak
rotation rate of a  galaxy to its intrinsic  luminosity.  In  the left
panel of Figure~8 we show NGC~1365 (in addition to NGC~925~(Silbermann
et~al.   1996),   NGC~4536~(Saha et~al.  1996)   and NGC~4639~(Sandage
et~al.  1996) added to   the ensemble of calibrators having  published
Cepheid distances from  ground-based data~(Freedman  1990), and I-band
magnitudes and  line widths,   measured  at 20\%  of  the peak  height
(Pierce  1994, Pierce  \&  Tully  1992 and  references therein).    As
mentioned earlier  NGC~1365 provides the  brightest data  point in the
relation;  additional  galaxies   recently  having Cepheid   distances
measured include  NGC~3621~(Rawson 1997), NGC~3351~(Graham  1997)  and
NGC~2090~(Phelps 1998),  and will be  included  once I-band magnitudes
become available.

Although we have only the Fornax cluster for comparison at the present
time, it is interesting  to note that  there is no obvious discrepancy
in  the Tully-Fisher relation  between  galaxies in the  (low-density)
field and  galaxies in this  (high-density) cluster  environment.  The
NGC~1365 data point   is consistent with   the data for  other Cepheid
calibrators.  Adding in all  of the other   Fornax galaxies for  which
there are published  I-band  magnitudes  and  inclination-corrected HI
line widths provides us with  another comparison of field and  cluster
spirals.  In  the right panel  of Figure~8 we  see  that the 21 Fornax
galaxies  (shifted by  the true  modulus  of NGC~1365) agree extremely
well with the 9 brightest Cepheid-based calibrators.  The slope of the
relation is virtually unchanged  by this augmentation. The  scatter in
the individually     Cepheid-calibrated   data  (left      panel)   is
$\pm$0.35~mag. This increases   to $\pm$0.48~mag if  the entire Fornax
cluster sample is  included (right panel).   In following applications
we  adopt  $M_I  = -8.80~(log(\Delta V)   - 2.445)  +   20.47$  as the
best-fitting  least squares solution (derived  from equal weighting of
all  galaxies and minimizing magnitude  residuals) for the calibrating
galaxies.

Han~(1992) has presented I-band  photometry and  neutral-hydrogen line
widths for the determination  of Tully-Fisher distances to  individual
galaxies in 16 clusters out to redshifts  exceeding 10,000~km/sec.  We
have rederived  distances and uncertainties  to each of these clusters
using  the above-calibrated expression  for the Tully-Fisher relation.
The  results are contained in  Figure~9.  A linear fit  to the data in
Figure~9 gives a Hubble constant of $H_o =$ 76~km/sec/Mpc with a total
observed scatter giving  a formal (random) uncertainty  on the mean of
only $\pm$2~km/sec/Mpc,  increasing  to $\pm$3   (Table  2) when  flow
uncertainties are added in quadrature.  It is significant that neither
Fornax nor  Virgo deviate to  any  significant degree  from an  inward
extrapolation of  this    far-field solution.  At  face   value, these
results provide evidence for both of  these clusters having only small
motions with  respect to their local Hubble  flow. This value compares
favorably with other recent  calibrations of the Tully-Fisher relation
by Giovanelli {\it et~al}  (1997) who obtain $H_o = 69\pm5$~km/sec/Mpc
(one      sigma)   and   then     by  Tully~(1998)    who   finds $H_o
= 82\pm16$~km/sec/Mpc (95\% confidence).

\medskip
\medskip
\centerline{\bf  10. OTHER  RELATIVE  DISTANCE DETERMINATIONS}   

In addition to the relative distances  using the Tully-Fisher relation
discussed above,  a set of relative  distance moduli based on a number
of  independent  secondary distance   indicators, including  brightest
cluster       galaxies,   Tully-Fisher   and   supernovae     is  also
available~(Jerjen \& Tammann  1993).  We adopt,  without modification,
their differential distance scale and tie into the Cepheid distance to
the  Fornax  cluster, which was  part  of their  cluster  sample.  The
results are  shown  in Figure~10 which extends   the velocity-distance
relation out to  more than 160~Mpc.  No  error  bars are  given in the
published  compilation.  For a  discussion   of uncertainties in  this
sample see  Huchra (1995).   This sample  yields  a  value of  $H_o =$
72~$(\pm3)_r$~km/sec/Mpc  (random),   with a  systematic  error of 9\%
being associated  with the  distance  (but not   the velocity) of  the
Fornax cluster.

\vfill\eject
\medskip
\medskip
\centerline{\bf  11. BEYOND  FORNAX:   TYPE  IA     SUPERNOVAE}

The Fornax cluster elliptical  galaxies NGC~1316 and NGC~1380 are host
to the well-observed type Ia supernovae 1980N and 1992A, respectively.
Although the distances to these  galaxies  are not measured  directly,
the new Cepheid  distance to NGC~1365,  and associated estimate of the
distance to the Fornax cluster allows two additional very high-quality
objects to be added to the calibration of type Ia supernovae, allowing
for the uncertainty in  their distances.  A preliminary discussion  of
these objects was given by Freedman et~al.   (1997).  A more extensive
discussion  of the  type Ia   supernovae  distance scale  (including a
re-analysis of  all of the   Cepheid data for type  Ia  supernova-host
galaxies) will be presented in Gibson et~al.  (1999, in preparation).

The galaxies hosting type   Ia supernovae for which Cepheid  distances
have been measured to date include:  IC~4182 (1937C), NGC~5253 (1895B,
1972E), NGC~4536   (1981B),  NGC~4496 (1960F),  NGC~4639  (1990N) (see
Sandage  et~al.   1996), and NGC~4414   (1974G) (Turner et~al.  1998).
NGC~3627 was host to 1989B, a galaxy which in the Leo Triplet, assumed
to be at the same distance as  the Leo I  Group (given by Cepheids) by
Sandage  et~al.  The  quality of the  supernova  observations for this
sample  is  quite  mixed; with   the exception of   1972E, the (mainly
photographic) photometry for the  earlier, historical supernovae is of
significantly lower quality than the more recent supernovae.

We have undertaken    a preliminary  recalibration  of   the type   Ia
supernovae, including SN~1980N and SN~1992A in the analysis. We assume
for  this purpose  that   the    Cepheid  distance  to  NGC~1365    is
representative of the  Fornax cluster, and  give these two  supernovae
only  half-weight compared to  the  other  objects  in the sample   to
reflect the additional uncertainty in  the distance.  For consistency,
we do this also for SN~1989B.  NGC~3627 was also host to SN~1973R, but
photographic  observations only are available  for this object.  Along
with SN~1895B, SN~1937C,  SN~1960F, and SN~1974G,  we currently do not
include the historical data  in this analysis. This procedure  differs
markedly from that of Sandage  et~al.  (1996), but is consistent  with
that of Hamuy et~al.  (1995, 1996).

Published   supernova magnitudes, errors,  and  decline rates plus the
Cepheid distances  are adopted from the  above sources.  Decline rates
and supernova magnitudes for  SN~1980N and SN~1992A were obtained from
Hamuy et~al.  (1991)  and Phillips (private communication). For 1980N,
we  adopt peak supernova  magnitudes of B  = 12.60$\pm$  0.03 mag, V =
12.44$\pm$0.03  mag, and $\delta  m_{15}$  =  1.28.  For  SN~1992A, we
adopt B   = 12.49$\pm$0.03 mag,  V =   12.55$\pm$0.03 mag and  $\delta
m_{15}$ = 1.47.    These    two supernovae are  amongst    the fastest
decliners    in  the  Cepheid-calibrating  sample.    A   decline-rate
absolute-magnitude   relation for the   Cepheid  calibrators  has been
presented by Freedman (1997).  It is consistent with that observed for
distant supernovae ({\it e.g.} Hamuy et~al.  1995, 1996).

The  Cepheid-calibrated supernova  sample  including the best-observed
supernovae   SN~1972E,  SN~1981B,  SN~1990N,   giving   half-weight to
SN~1989B, SN~1980N  and SN~1992A, and  applied to the  distant Type Ia
supernovae of Hamuy (1995) gives  H$_o$ = 67~km/sec/Mpc.  We have also
experimented  with  various  weighting  schemes  (e.g., including  the
photographic  data;  while   allowing  for  its   larger  uncertainty;
excluding the Fornax data  completely; or, analyzing the  B and V data
alone).   Resulting values  of   H$_o$ lie in     the range of   63-67
km/sec/Mpc.  Half of  the  difference between  the  results of Sandage
et~al.   (1996) (giving H$_o$ =   57 km/sec/Mpc) is  the lower  weight
placed  in the current analysis on   the (poorer quality) photographic
data.  The remaining  difference is due to  our adoption of the  Hamuy
et~al.  (1995) decline-rate   correction,  and the  inclusion   of the
Fornax supernovae.  Sandage  et~al.  adopt no decline-rate correction.
The   relation that we   have  adopted  is   consistent with   that of
Phillips~(1993),   Hamuy  et~al.~(1996)    and    Reiss,   Press    \&
Kirshner~(1996).   Finally, we note  that  eliminating the new  Fornax
calibrators from the  analysis changes the  Hubble constant  by $-$1-3
km/sec/Mpc,  for a  variety    of calibrating samples  and   weighting
schemes.

\vfill\eject
\medskip
\medskip
\centerline{\bf  12. COMPARING AND COMBINING THE RESULTS}

The results of  the previous five  sections are presented  in Table 3.
What  is the summary conclusion?  In  the first instance we can simply
state that  based on a  number  of different methods  calibrated here,
$H_o$     falls   within     the   range      of   67$(\pm6)_r$    and
76$(\pm3)_r$~km/sec/Mpc, with no obvious  dependence on the indicative
volume  of  space being  probed.   Hence,   a  variety  of independent
distance determination   methods are  yielding agreement  at  the 10\%
level.  With   the exception of  the common  Cepheid PL  relation zero
point, these various   determinations  are largely independent;   thus
their  differences  are  indicative  of  the  true   systematic errors
affecting   each   of the  methods  and    their individual underlying
assumptions.  No  single  determination stands out as  either markedly
anomalous or as undeniably superior.

How then do  we combine these individual  results in a summary  number
with its own uncertainty? We have undertaken  two types of approach: a
{\it Frequentist} approach and a {\it Bayesian}  one.  In the end they
only  differ  in their resulting   confidence  intervals.  We begin by
first considering the random errors.

In our application of the Frequentist  approach ({\it e.g.,} Wall 1997
and references  therein) we simply represent  each determination  as a
probability distribution having its  mean  at $H(i)$, a dispersion  of
$\sigma(i)_{random}$ and unit integral ({\it i.e.,} equal total weight
in the  sum).  These are  shown as the connected  dotted lines  in the
left panel of Figure~11.   The solid enveloping  line is the resulting
sum  of the five  probability  density  distributions.  The  composite
probability  distribution is  somewhat non-Gaussian,  but it is  still
centrally peaked  with  both the  mode and   the median  coinciding at
72-73~km/sec/Mpc.    An estimate  of the  traditional ($\pm$one-sigma)
errors can be easily obtained   from this distribution by  identifying
where  the cumulative probabilities   hit 0.16 and 0.84, respectively.
This procedure gives the cited error on the mean for five estimates of
$\pm5$~km/sec/Mpc. [At the suggestion of the referee this exercise was
repeated  using  identical  errors of   10\%    on each  of  the  five
estimates. The result is 71$\pm$4~km/sec/Mpc.]

The  Bayesian estimate is   equally straightforward (see  Press 1997).
Again,   taking  the  individual  Hubble    constant estimates to   be
represented by Gaussians we combine them by multiplication, assuming a
minimum-bias  (flat) prior (Sivia   1996).  We note that  the Bayesian
approach assumes statistical  independence; and  strictly speaking the
considered   samples are not   completely  independent given that they
explicitly share  a common  (Cepheid)  zero point,  and the Jerjen  \&
Tammann  hybrid  sample overlaps in part    with the pure Tully-Fisher
application.  Nevertheless we have done  the exercise of computing the
posterior probability distribution  with these caveats clearly stated.
Because of  the strong overlap  in the various estimates  the combined
solution is both very strongly peaked  and symmetric giving a value of
$H_o =   74 (\pm3)_r$~km/sec/Mpc as   depicted in the   right panel of
Figure~11.  The one-sigma error on the  mean was again determined from
the 0.16  and  0.84 cumulative  probability distribution   points.  As
already   anticipated  above the  results  of   the two  analyses  are
indistinguishable except for   the high confidence   attributed to the
number by the Bayesian    analysis.   These results are     summarized
graphically in Figure~11 and numerically in Table~3.

Systematic errors must be dealt with separately and independently from
the random error discussion.  While  some of the identified systematic
errors affect all of the above determinations  equally and in the same
sense   (the  LMC distance modulus   for   instance), others  are more
`randomly' distributed   among  the methods    and their  contributing
galaxies.  For  instance, the  (as yet unknown)   effects of flows are
estimated and scaled  for each of  the  methods here, but  they may be
large for the Fornax cluster, but smaller and perhaps have a different
sign for the ensemble   of   type Ia  supernovae.  It   seems  prudent
therefore to simply average   the systematic errors while listing  out
the main components individually.   They main systematic error on  the
finally adopted value of the Hubble constant are: (1) large scale flow
fields. These contribute large fractional uncertainties to the nearest
estimates of H$_o$, but they progressively  drop to only a few percent
at large  distances and/or for samples   averaged over many directions
(see Table 2 and further discussion below).  (2) The zero point of the
adopted PL relation. In our case, this is tied directly to the adopted
true distance modulus of the LMC.   A variety of independent estimates
are reviewed by Westerlund (1996) and  more recently by Walker (1999);
they  each conclude that   the  uncertainty is at  the  5\%  level  in
distance.    Westerlund     prefers a   true    distance   modulus  of
18.45$\pm0.10$mag,      Walker    adopts 18.55$\pm0.10$mag.   We  have
consistently used 18.50$\pm0.10$mag throughout  this series of papers.
Therefore,  if   the   distance  to the   LMC  systematically  changes
downward/upward (by 10\% or 0.2~mag, say) from that adopted here, then
our entire distance  scale shifts by  the same amount, and the derived
value of  $H_o$  would  increase/decrease by the  same  (10\%) factor.
And, finally,  (3) the  metallicity    dependence of  the Cepheid   PL
relation.  This is a complex and much debated topic and the interested
reader  is referred to    Sasselov  et~al.  (1997),  Kochanek  (1997),
Kennicutt et~al.  (1998),  and earlier reviews  by Freedman \&  Madore
(1990) for an introduction to  the literature.  The metallicity of the
galaxies for which Cepheid searches  have been undertaken span a range
in [O/H] abundance   of almost an order   of magnitude, with  a median
value  of --0.3 dex.  The  calibrating  sample of Cepheids  in the LMC
have a very  similar  abundance of  [O/H] = --0.4  dex.  These results
suggest that even if in    individual  cases the metallicity    effect
amounted to   10-20\%,   the overall   effect  on  the  calibration of
secondary distance    indicators  will be  less  than   5\%.  Recently
concluded  observations  with NICMOS on   HST  should further  help to
constrain the magnitude of this effect.

The importance  of bulk flow  motions on  the  determination of  H$_0$
varies  significantly depending  on   how far   a  particular distance
indicator  can  be extended  to  (see  Table 2).   For  local distance
indicators   the uncertainties due to    unknown bulk motions are  the
largest contributing source of  systematic error to the  determination
of H$_o$, amounting to an uncertainty of 20-25\% in the local value of
the Hubble constant.  The Tully-Fisher  relation extends to a velocity
distance of    about 10,000 km/sec,   although most  of   the observed
clusters  are not this remote.  At  6,000 km/sec,  peculiar motions of
$\sim$300  km/sec   would individually contribute  5\%  perturbations;
however, with many clusters distributed over the sky, peculiar motions
of this magnitude will give only a few  percent uncertainty.  For type
Ia supernovae  which extend  to  beyond 30,000  km/sec, the problem is
even less severe.  These estimates are  consistent with recent studies
by Shi \&  Turner (1998) and Zehavi et~al.   (1998) which place limits
on the variation of H$_o$   with  distance, based on theoretical   and
empirical considerations, respectively.

\medskip
\medskip
\centerline{\bf  14. COSMOLOGICAL  IMPLICATIONS}

A  value of the  Hubble constant, in  combination  with an independent
estimate of the  average  density of the    Universe, can be  used  to
estimate a dynamical age for the Universe ({\it e.g.}, see Figure~12).
For a value  of of  $H_o$ =  72~$(\pm5)_r$~km/sec/Mpc, the age  ranges
from a  high  of $\sim$12  Gyr  for  a low-density  ($\Omega =  0.15$)
Universe,  to  a  young age of   $\sim$9   Gyr for  a critical-density
($\Omega  =  1.0$) Universe.  These  ages  change to  15 and  7.5 Gyr,
respectively allowing for an error of $\pm10$ km/sec/Mpc.

The ages  of Galactic globular  clusters have until recently tended to
fall in the range    of 14$\pm$2 Gyr  (Chaboyer, Demarque,   Kernan \&
Krauss   1996); however,  the  subdwarf  parallaxes  obtained  by  the
Hipparcos  satellite~({\it  e.g.} Reid   1997) may  reduce  these ages
considerably.   For $\tau =$  14~Gyr  and $\Omega  = 1.0$, $H_o$ would
have to be $\sim$45~km/sec/Mpc; interpreted within  the context of the
standard Einstein-de~Sitter model, our value of $H_o$ = 72 km/sec/Mpc,
is  incompatible with a  high-density ($\Omega =  1.0$) model universe
without a cosmological constant  (at the 2-sigma  level defined by the
identified systematic errors.) If, however, $\tau  =$ 11~Gyr, then the
globular cluster and the expansion ages would  be consistent to within
their mutually quoted uncertainties.

\vfill\eject
\medskip
\medskip
\medskip
\centerline{\bf Acknowledgements} 

This  research  was supported by  the   National Aeronautics and Space
Administration (NASA) and the  National Science Foundation (NSF),  and
benefited from the use of the  NASA/IPAC Extragalactic Database (NED).
Observations  are based on data  obtained  using the {\it Hubble Space
Telescope} which is operated  by the Space Telescope Science Institute
under contract  from the Association  of Universities for  Research in
Astronomy. LF acknowledges support by  NASA through Hubble  Fellowship
grant   HF-01081.01-96A awarded  by   the   Space   Telescope  Science
Institute, which  is operated by the  Association of  Universities for
Research in  Astronomy, Inc., for NASA under  contract NAS 5-26555. We
thank the  referee, George  Jacoby, and  the Editor,  Greg Bothun, for
numerous detailed and insightful comments on the paper.

\medskip
\medskip
\medskip
\medskip
\medskip
\centerline{\bf REFERENCES} 

\par\noindent
Aaronson, M., Mould, J., Huchra, J., Sullivan, W.~T., Schommer, R.~A.,
\& Bothun, G.~D. 1980, ApJ, 239, 12

\par\noindent
Bureau,  M., Mould, J.~R., \&  Staveley-Smith,  L. 1996 ApJ, 463,  60,
1996

\par\noindent
Chaboyer, B.,  Demarque, P.,  Kernan, P.~J.,  \&  Krauss, L.~M.  1996,
Science, 271, 957

\par\noindent
Coles, P., \& Lucchin, F., 1995, in {\it Cosmology,} Wiley, 399

\par\noindent
de Vaucouleurs, G., 1975 in {\it Stars and  Stellar Systems,} {\bf 9},
eds.    A.~R.~Sandage,  M.~Sandage,  J.~Kristian, Univ  Chicago Press:
Chicago, p.~557

\par\noindent
Drinkwater,   M.~J., \&   Gregg,    M.~D. 1998,  MNRAS, 296, L15

\par\noindent
Feast, M.~W., \& Catchpole, R.~M. 1997, MNRAS, 286, L1

\par\noindent
Ferguson, H.~C. 1989, AJ,  98, 367

\par\noindent
Ferguson,  H.~C.,  \& Sandage,  A.~R. 1988, AJ,  96, 1520

\par\noindent
Ferrarese, L.  et~al. 1996, ApJ, 464, 568

\par\noindent
Freedman,  W.~L.  1990, ApJL, 355, L35

\par\noindent
Freedman, W.~L., \&   Madore, B.~F.  1990, ApJ, 365, 186

\par\noindent
Freedman, W. L. 1997, in ``Critical Dialogs in Cosmology'', ed. N. Turok, 

Princeton 250th Anniversary Conference, June 1996, (World Scientific), 
pp. 92-129.

\par\noindent
Freedman,  W.~L., Madore, B.~F.  \& Kennicutt,  R.  1997,  in {\it The
Extragalactic    Distance   Scale},   eds.   M.~Livio,  M.~Donahue  \&
N.~Panagia, Cambridge Univ. Press: Cambridge, 171

\par\noindent
Freedman, W.~L., et~al. 1994a, ApJ, 435, L31

\par\noindent
Freedman, W.~L., et~al. 1994b, Nature,  371, 757,

\par\noindent
Freedman, W.~L., et~al. 1998, in preparation

\par\noindent
Giovanelli, R., Haynes, M.~P., Da Costa, L.~N., Freudling, W., Salzer,
J.~J., \& Wegner, G. 1997, ApJ, 477, L1

\par\noindent
Graham, J.~A.,  et~al. 1997, ApJ,  477, 535

\par\noindent
Hamuy, M., et~al. 1991, AJ, 102, 208 

\par\noindent
Hamuy, M., et~al. 1995, AJ, 109, 1

\par\noindent
Han, M.~S.  1992,  ApJS,  81, 35

\par\noindent
Han M., \& Mould, J.~R. 1990, ApJ, 360, 448

\par\noindent
Hill, R.,  et~al. 1998, ApJ, 496, 648

\par\noindent
Holtzmann, J., et~al. 1995, PASP, 107, 156

\par\noindent
Hubble, E.~P. 1929,   Proc. Nat. Acad. Sci.,  15, 168

\par\noindent
Huchra, J.,  Geller, M., Clemens, C., Tokarz, S.  \& Michel, A. 1992,
Bull. CDS, 41, 31.

\par\noindent
Huchra, J.  1995, Heron  Island Conference on {\it Peculiar Velocities
in the Universe}, P. Quinn and W.  Zurek, eds. (published on the World
Wide Web at http://msowww.anu.edu.au/~heron).

\par\noindent
Jacoby, G., et~al. 1992, PASP, 104, 599 

\par\noindent
Jensen, J.B., Tonry, J.L., \& Luppino, G.A. 1998, ApJ, 505, 111

\par\noindent
Jerjen, H., Freeman, K.C., \& Binggeli, B. 1998, AJ, in press

\par\noindent
Jerjen, H., \& Tammann, G.~A. 1993, A\&A, 276, 1

\par\noindent
Kennicutt, R.~C., Freedman, W.~L., \& Mould, J.~R. 1995, AJ, 110, 1476

\par\noindent
Kennicutt, R.~C., et~al. 1998, ApJ, 498, 181

\par\noindent
Kochanek, C.~S.  1997, ApJ, 491, 13

\par\noindent
Kraan-Korteweg, R. 1986, A\&AS, 66, 255

\par\noindent
Lafler, J., \& Kinman, T.~D. 1965, ApJS, 11, 216

\par\noindent
Lauer, T.D., Tonry,  J.L.,  Postman, M.,  Ajhar, E.A.,  \&  Holtzmann,
J.A. 1998, ApJ., 499, 577

\par\noindent
Madore,  B.~F.,  Freedman,  W.~L.  \&   Sakai, S.   1997,  in {\it The
Extragalactic  Distance Scale},    eds.    M.~Livio,  M.~Donahue    \&
N.~Panagia, Cambridge Univ. Press: Cambridge, 239

\par\noindent
Madore, B.~F., et~al. 1998, Nature, 395, 47

\par\noindent
Madore, B.~F., \&  Freedman, W.~L. 1991, PASP, 103, 933

\par\noindent
Madore, B.~F., \&  Freedman, W.~L. 1998,  ApJ, 492, 110

\par\noindent
Mathewson, D.~S., Ford, V.~L., \& Buchhorn, M. 1992, ApJS, 81, 413

\par\noindent
McMillan, R., Ciardullo, R., \& Jacoby, G.H. 1996, ApJ, 416, 62

\par\noindent
Mould, J.~R.,  Sakai,  S., Hughes, S., \&  Han,  M. 1997,  in {\it The
Extragalactic    Distance   Scale},   eds.   M.~Livio,  M.~Donahue  \&
N.~Panagia, Cambridge Univ. Press: Cambridge, 158

\par\noindent
Mould, J.~R., et~al. 1995, ApJ, 449, 413

\par\noindent
Phelps, R.,  et~al. 1998, ApJ, 500, 763

\par\noindent
Pierce, M. 1994,  ApJ, 430, 53

\par\noindent
Pierce, M., \& Tully, R.B. 1992,  ApJ, 387, 47

\par\noindent
Press,   W. 1997, in   {\it  Unsolved Problems  in Astrophysics},  eds
J.P.~Ostriker \& J.N.~Bahcall, Princeton Univ. Press: Princeton, p.49

\par\noindent
Rawson, D.~M.,  et~al. 1997,  ApJ, 490, 517

\par\noindent
Reid, N. AJ, 114, 161, 1997

\par\noindent
Riess, A.~C., Press, W.~H. \& Kirschner, R.~P. 1996, ApJ, 473, 88 

\par\noindent
Saha, A., Sandage, A.~R., Labhardt, L., Tammann, G.~A., Macchetto,
F.~D., \& Panagia, N. 1996, ApJ, 466, 55

\par\noindent
Sandage, A.~R., Saha, A., Tammann, G.~A., Labhardt,  L., Panagia, N., \&
Macchetto, F.~D. 1996, ApJL, 460, L15

\par\noindent
Sasselov, D.~D., et~al. 1997, A\&A, 324, 471

\par\noindent
Schechter, P. 1980, AJ,  85, 801

\par\noindent
Schechter, P., Mateo, M. Saha, A. 1993, PASP,  105, 1342

\par\noindent
Schroder, A. 1995, ~Doctoral Thesis, University of Basel

\par\noindent
Shi, X., \& Turner, M. 1998, ApJ, 493, 519

\par\noindent
Silbermann, N.~A., et~al.  1996, ApJ, 470, 1

\par\noindent
Silbermann, N.~A., et~al. 1999, ApJ, in press

\par\noindent
Sivia, D.~S., 1996, Data Analysis: A Bayesian Tutorial, Claredon Press:
Oxford

\par\noindent
Stetson, P.~B. 1994, PASP,  106, 250

\par\noindent
Tammann,  G.~A.  \& Federspiel,   M.  1997, in {\it  The  Extragalactic
Distance Scale}, eds.  M.~Livio, M.~Donahue \& N.~Panagia, Cambridge
Univ. Press: Cambridge, 137

\par\noindent
Tully, R.~B.  1998, in {\it  Cosmological Parameters of the Universe},
IAU Symp. 183, ed.  K.~Sato, Reidel: Dordrecht, (in press)

\par\noindent
Turner, A., et~al. 1998, ApJ, 505, 87

\par\noindent
Wall, J.~V.  1997, QJRAS, 37, 519

\par\noindent
Westerlund, B.  1996, in {\it  The  Magellanic Clouds,} Cambridge
Univ. Press: Cambridge

\par\noindent
Zehavi, I, Reiss, A., Kirshner, R.~P., \& Dekel, A. 1998, ApJ submitted
astrop-ph 9802252

\vfill\eject
 
\centerline{\bf FIGURE CAPTIONS}
\par\noindent
{\bf Fig.~1.\ \ --  } A comparison of the  distribution of galaxies as
projected on   the sky for the  Virgo  cluster  (right  panel) and the
Fornax cluster (left  panel). M100 and  NGC~1365 are each individually
marked by  arrows showing their relative   disposition with respect to
the  main body  and cores of   their  respective clusters.  Units  are
arcmin.

\par\noindent
{\bf   Fig.~2.   \  \ --    } Fornax  galaxies with  published  radial
velocities  within 6\deg~ of  the  cluster center and having  apparent
velocities  less than 2,500~km/sec.  All 117   galaxies used to define
the mean velocity (and velocity dispersion) for the Fornax cluster are
shown  plotted as they appear on  the sky.  The 78 early-type galaxies
are depicted by filled circles; the 39 late-type  galaxies are show as
open  circles.  NGC~1365,   near the   center   of  the cluster,    is
individually marked.

\par\noindent
{\bf Fig.~3. \ \ -- } Velocity-position plots for 117 Fornax galaxies.
The  right-hand portion of the  figure shows the galaxies projected in
declination down onto a right ascension slice of the sky.  NGC~1365 is
marked, and seen     to be centrally   located in   both  position ans
velocity.  Open  circles are spiral/irregular galaxies; filled circles
represent E/S0 galaxies.

\par\noindent
{\bf Fig.~4.  \ \  -- } The  velocity structure of the Fornax cluster.
The upper  left insert shows that  simple binned  histogram of the 117
velocities for galaxies in  the  Fornax cluster.  The distribution  is
symmetric about 1,400~km/sec and is closely approximated by a Gaussian
with a one-sigma width of  $\sim$340~km/sec. Another representation of
the velocity density distribution is given in the  main portion of the
left  panel.  Here we have   represented each galaxy as an  individual
Gaussian of unit weight centered at its quoted velocity and widened by
its   published  uncertainty  (tall  spikes   represent high-precision
velocities; low, broad smears  represent uncertain observations).  The
solid curve is the sum of the individual Gaussians.

To obtain  a mean  and sigma from  this probabilistic  distribution we
refer to  the right  panel where  the cumulative  probability  density
(CPD) distribution is plotted. Horizontal   lines at CPD = 0.50,  0.16
and 0.84 cross  the  distribution curve at  the mean  velocity and  at
$\pm$one-sigma, respectively.  These are  to be compared to the simple
average and  standard deviation shown by  the  centrally plotted error
bar. The close coincidence of the two estimates is a direct reflection
of the highly Gaussian nature  of the Fornax velocity distribution. At
the base  of each of  the plots  the  velocity  of NGC~1365 is  shown,
fitting well within the one-sigma velocity dispersion.

\par\noindent
{\bf Fig.~5.   \ \ -- } V  and I-band  Period-Luminosity relations for
the 37 Cepheids discovered in NGC~1365.   The fits are to the fiducial
relations given  by (Madore \& Freedman  1991) shifted to the apparent
distance  modulus of   NGC~1365.  Dashed lines  indicate the  expected
intrinsic ($\pm$2-sigma) width of  the relationship due to  the finite
temperature width of the Cepheid instability strip.  The solid line is
a minimum $\chi  ^2$ fit to the fiducial  PL relation for LMC Cepheids
(18), corrected for $E(B-V)_{LMC}  =$ 0.10 mag,  scaled to an LMC true
distance  modulus  of   $\mu _o$    =   18.50 mag,  and   shifted into
registration with  the  Fornax data.  [Note:  Recent  results from the
Hipparcos satellite bearing on the Galactic calibration of the Cepheid
zero point (Feast \& Catchpole  1997; Madore\& Freedman 1998) indicate
that the  LMC calibration is  confirmed  at the level  of  uncertainty
indicated in  Table  1, with   the possibility that  a small  (upward)
correction to the LMC reddening may be indicated.]

\par\noindent
{\bf Fig.~6.   \ \   -- }  Relative  geometry (left  panel), and   the
corresponding velocity vectors (right  panel) for the disposition  and
flow of Fornax  and   the Local  Group  with   respect to the    Virgo
cluster. The circles plotted at the  positions of the Virgo and Fornax
clusters have the same angular size as  the circles enclosing M100 and
NGC~1365 in the two panels of Figure~1.

\par\noindent
{\bf   Fig.~7.  \  \  --  }  The velocity-distance  relation for local
galaxies  having Cepheid-based   distances.    Circled dots mark   the
velocities   and distances of  the   parent groups  or  clusters.  The
one-sided  ``error''   bars  with  galaxy   names  attached   mark the
velocities associated with   the  individual galaxies  having   direct
Cepheid distances. The heavy broken line  represents a fit to the data
giving   $H_o    =    73$~km/sec/Mpc.  The    observed    scatter   is
$\pm$12~km/sec/Mpc, and is shown by the thin diverging broken lines.

\par\noindent
{\bf Fig.~8.  \ \ --  } Tully-Fisher relations.   The left panel shows
the absolute  I-band magnitude, $M_I$ versus the inclination-corrected
21-cm line widths  (measured at 20\%  of the peak) for galaxies having
individually determined Cepheid  distances.  NGC~1365 is the brightest
object  in this sample; the position  of  this {\it cluster spiral} is
consistent with an extrapolation of  the relation defined by the lower
luminosity {\it field  galaxy sample.}   The   right panel  shows  the
calibrating  sample   (filled circles)    superimposed on  the  entire
population of Fornax spiral galaxies for which I-band observations and
line widths are available (Bureau, Mould  \& Staveley-Smith 1996); the
latter being shifted to absolute magnitudes by the Cepheid distance to
NGC~1365. No errors are tabulated   for the field galaxy  calibrators;
error  bars for the Fornax sample  are as given  in Table 1 in Bureau,
Mould \& Staveley-Smith.

\par\noindent
{\bf Fig.~9. \ \ -- } The  velocity--distance relation for 16 clusters
of galaxies out to  11,000~km/sec,  having distance moduli  determined
from  the  I-band Tully-Fisher relation.   A  fit to  the data gives a
Hubble  constant  of  $H_o = 76$~km/sec/Mpc.    The   solid lines mark
one-sigma   bounds on the  observed  internal  scatter.  The range  of
distance and  velocity probed directly  by Cepheids, as illustrated in
Figure 7, is outlined at the bottom left corner of this figure.

\par\noindent
{\bf Fig.~10. \ \ -- } The velocity--distance relation for 17 clusters
of galaxies,  having  published (Jerjen \& Tammann  1993) differential
distance moduli scaled to the Fornax cluster.  A fit to the data gives
a Hubble constant of $H_o = $72~km/sec/Mpc.  As in Figure~7, the solid
lines mark one-sigma bounds on the observed internal scatter.

\noindent
{\bf Fig.~11.  \ \ -- } A graphical representation of Table 3 is given
in the  left panel, showing the various   determinations of the Hubble
constant,   and the  adopted  mean.    Each  value of   $H_o$ and  its
statistical uncertainty is  represented  by a  Gaussian of unit   area
(linked dotted  line) centered on  its determined  value  and having a
dispersion equal  to   the quoted  {\it  random}  error.    Superposed
immediately above each  Gaussian is a  horizontal bar representing the
one sigma limits of the calculated {\it systematic errors} derived for
that determination.    The adopted average  value and  its probability
distribution function (continuous solid line) is the arithmetic sum of
the  individual  Gaussians.  (This  Frequentist representation  treats
each determination  as  independent,  and  assumes no  {\it a  priori}
reason   to  prefer   one   solution  over    another.)   A   Bayesian
representation  of the products   of  the various probability  density
distributions is  shown  in the right   panel.  Because of   the close
proximity and strong overlap in the  various independent solutions the
Bayesian estimator  is  very similar to,  while  more sharply  defined
than, the Frequentist solution.

\noindent
{\bf  Fig.~12.   \ \  --   }  Lines  of  fixed  time representing  the
theoretical ages  of the oldest globular  cluster stars are  shown for
12, 14 and 16 Gyr, plotted  as a function  of the expansion rate $H_o$
and density parameter $\Omega_o$,  for an  Einstein-de~Sitter universe
with the    cosmological constant $\Lambda  =   0$.  The  thick dashed
horizontal line  at  $H  =  72~(\pm5)_r  ~[\pm7]_s$~km/sec/Mpc  is the
average value of  the Hubble constant  given in Table 3.  The parallel
(solid) lines on either side of  that solution represent the one-sigma
random errors on that solution.  Systematic errors on the solution for
$H_0$ are  represented by thin dashed  lines at 65  and 79~km/sec/Mpc.
The only region of (marginal) overlap between these two constraints is
in the low density ($\Omega <$ 0.2) regime, unless $\Lambda \ne 0.$ If
the globular cluster ages are assumed to place {\it  a lower bound} on
the age  of the Universe, the  region of plausible overlap between the
two  solutions  is  more  severely restricted  to even   lower density
models.

\vfill \bye